\documentclass[pra,twocolumn,showpacs,preprintnumbers,superscriptaddress]{revtex4}

\usepackage{amsmath,amsfonts,amssymb,bbm,slashed}
\usepackage[english]{babel}
\usepackage{graphicx}
\usepackage{color}

\begin{document}

\title{The ``Majoranon'' and how to realize it in a tabletop experiment}

\author{Changsuk Noh}
\email{cqtncs@nus.edu.sg}

\affiliation{Centre for Quantum Technologies, National University of Singapore, 2 Science Drive 3, Singapore 117542.}
\author{B. M. Rodr\'{\i}guez-Lara}
\affiliation{Centre for Quantum Technologies, National University of Singapore, 2 Science Drive 3, Singapore 117542.}
\affiliation{Instituto Nacional de Astrof\'{i}sica, \'Optica y Electr\'{o}nica \\
Calle Luis Enrique Erro No.~1, Sta. Ma. Tonantzintla, Pue. CP 72840, M\'{e}xico}

\author{Dimitris G. Angelakis}
\email{dimitris.angelakis@gmail.com}
\affiliation{Science Department, Technical University of Crete, Chania, Crete, Greece, 73100}
\affiliation{Centre for Quantum Technologies, National University of Singapore, 2 Science Drive 3, Singapore 117542.}
\begin{abstract}
We introduce the term Majoranon to describe particles that obey the Majorana equation, which are different from the Majorana fermions widely studied in various physical systems. A general procedure to simulate the corresponding Majoranon dynamics, based on a decomposition of the Majorana equation into two Dirac equations, is described in detail. It allows the simulation of the two-component chiral spinors, the building blocks of modern gauge theories, in the laboratory with current technology. Specifically, a Majoranon in one spatial dimension can be simulated with a single qubit plus a continuous degree of freedom, for example a single trapped ion. Interestingly, the dynamics of a Majoranon deviates most clearly from that of a Dirac particle in the rest frame, in which the continuous variable is redundant, making a possible laboratory implementation feasible with existing set ups.

\end{abstract}


\maketitle
\textbf{Introduction} -
Quantum simulation was originally envisioned as an approach to study complex quantum systems that are difficult to understand using conventional methods \cite{Feynman, Lloyd}.  However, it has recently been realized that the concept can also be used to engineer quantum dynamics not readily accessible in naturally occurring physical systems; e.g., elementary particles in lower spacetime dimensions. A notable example is an experimental demonstration of quantum simulation of the Dirac equation \cite{Dirac,Lamata} in 1+1 dimensional spacetime using trapped ions \cite{Gerritsma10}.  There have been other proposals and a few experimental demonstrations regarding the quantum simulation of relativistic equations and phenomena \cite{Gerritsma11,Casanova10,Noh}, including a recent proposal to simulate the Majorana equation \cite{Casanova}. 

The Majorana equation is a Lorentz covariant equation discovered by Majorana as a generalization of the Dirac equation \cite{Majorana1, Majorana2}:
\begin{eqnarray} \label{eq:MajoranaEquation}
\imath \gamma^\mu \partial_\mu \psi = m\psi_c,
\end{eqnarray}
where we have set $\hbar = c = 1$ for convenience. The gamma matrices $\gamma^\mu$ obey $\{ \gamma^\mu,\gamma^\nu\} = 2g^{\mu\nu}$; the symbol $\psi_c$ stands for charge conjugation of the spinor $\psi$: $\psi_{c} \equiv C (\gamma^0)^T  \psi^*$, where $C$ obeys $C(\gamma^\mu)^TC^{-1} = -\gamma^\mu$, $C^\dagger = C^{-1}$ and $C^T = -C$; $g^{\mu\nu}$ is the standard metric in the vacuum \cite{Giunti}. To avoid confusion, let us discuss here the nomenclature found in the literature. A field obeying the Majorana condition, $\psi_c = \psi$, is called a Majorana field (fermion, particle), in which case Eq.~(\ref{eq:MajoranaEquation}) reduces to the Dirac equation. The equation itself is sometimes called the Majorana equation when the field obeys the condition \cite{Wilczek}, but we will reserve the term Majorana equation for Eq.~(\ref{eq:MajoranaEquation}) when the condition does not hold and call the hypothetical particles that obey it Majoranons. We would like to emphasize that Majoranons are thus different to Majorana fermions that are widely studied in the literature.

The Majorana equation has received a renewed interest since the discovery of finite masses of neutrinos. A neutrino is a neutral fermion; therefore it could be either a Majorana fermion, in which case it is its own antiparticle, or a Dirac particle, in which case a neutrino will be different to an anti-neutrino. For this reason, most studies on Eq.~\eqref{eq:MajoranaEquation} are limited to Majorana fermions and the Majorana equation by itself seems to be of academic interest only. However, Casanova et al.~have recently proposed a general procedure to implement non-physical operations such as charge conjugation and time reversal, allowing for an experimental study of the equation \cite{Casanova}. 

Closely related to the Majorana equation are the so-called two-component Majorana equations (TCMEs). In 1+1 and 2+1 D, it turns out that one of the two TCMEs is equivalent to the Majorana equation, whereas in 3+1 D both of them merge into the Majorana equation. The importance of the TCMEs is that they are obeyed by the building blocks of the modern gauge theory, chiral spinors \cite{Aste, Pal}. In this work, we show that it is possible to decompose one of the equations, say for a right-chiral field, into two Hamiltonian equations that can be simulated. Our method differs from the earlier proposal \cite{Casanova}, which depends on enlarging the Hilbert space of the system to turn the Majorana equation into a higher-dimensional Dirac equation. An advantage is that the size of the Hilbert space required is smaller, but a tradeoff exists, in that a complete reconstruction of the spinor, including the continuous mode, is required. These points are discussed further below.

This paper is organized as follows. We start by giving a short summary on quantum simulation of the Majorana equation based on Hilbert space expansion mentioned above, followed by a brief introduction to TCMEs for non-specialists. We show how the equations can be motivated by trying to introduce Lorentz invariant mass terms in the Weyl equations \cite{Aste}, and describe how relativistic equations for higher-dimensional spinors can be constructed from these equations.  We then prove that a general fermion obeying what we will call a Dirac-Majorana equation can be decomposed into two Majorana particles \cite{Cheng,Giunti,Aste}, which implies that such an equation can be physically realized. We end by describing a general simulation procedure based on this decomposition.

\textbf{Quantum simulation by Hilbert space expansion} -
As shown in \cite{Casanova}, it is possible to implement unphysical operations such as charge conjugation, complex conjugation, and time reversal, via mapping to an enlarged Hilbert space. It is thus possible to simulate the Majorana equation which in 1+1 D reads
\begin{eqnarray}
\imath \partial_{t} \psi =  \sigma_{x} p_{x} \psi - \imath m \sigma_{y} \psi^*,
\end{eqnarray}
where $\psi = (\psi^{(1)} , \psi^{(2)})$, $\psi^{(i)} \in \mathbb{C}$ so $\psi \in \mathbb{C}^2$. To see how, first note that this equation can be rewritten as
\begin{eqnarray}
\imath \partial_{t} \left( \psi + \psi^{\ast} \right) =  \sigma_{x} p_{x} \left( \psi + \psi^{\ast} \right) + \imath m \sigma_{y} \left( \psi - \psi^{\ast} \right),\\
\imath \partial_{t} \left( \psi - \psi^{\ast} \right) =  \sigma_{x} p_{x} \left( \psi - \psi^{\ast} \right) - \imath m \sigma_{y} \left( \psi + \psi^{\ast} \right). 
\end{eqnarray}
Now, by mapping $\psi$ to an extended Hilbert space such that $\Psi = ( \mathrm{Re}(\psi^{(1)}),\mathrm{Re}(\psi^{(2)}),\mathrm{Im}(\psi^{(1)}),\mathrm{Im}(\psi^{(2)})  )$, $\psi = M \Psi$, $M = ( 1_{2} , \imath 1_{2})$, the Majorana equation can be written as the Schr\"odinger equation, $\imath \partial_{t} \Psi = H_{M} \Psi$, where
\begin{eqnarray}
\label{MHam}
H_{M} =   \left( 1_{2} \otimes \sigma_{x} \right) p_{x} - m (\sigma_{x} \otimes \sigma_{y} ).
\end{eqnarray}
This Hamiltonian, describing the Dirac equation, can be implemented with two trapped ions. 

\textbf{Two-component formalism} -
Here we give a brief introduction to the two component formalism of chiral spinors and show how they are related to the usual four-component Dirac equation and the Majorana equation \cite{Aste, Giunti, Pal}. 
Being the smallest irreducible representations of the Lorentz group, the chiral (or Weyl) spinors are the basic building blocks of modern gauge theories \cite{Giunti}. Moreover, chirality plays an important role in the Standard Model due to the chiral nature of electroweak interaction, i.e.~the electroweak interaction distinguishes between left- and right-chiral fields. 

Let us start with the description of massless fermions by the Weyl equations:
\begin{align}
(\partial_t-\vec{\sigma}\cdot\vec{\nabla})\psi_L = 0, \\ 
(\partial_t+\vec{\sigma}\cdot\vec{\nabla})\psi_R = 0,
\end{align}
where $\vec{\sigma} = (\sigma_x,\sigma_y,\sigma_z)$ are the Pauli matrices. The Weyl spinors $\psi_R$ and $\psi_L$ transform according to the two dimensional irreducible representation of the Lorentz group and its complex conjugate, respectively, and are called two-component chiral spinors. In the massless case, the chirality coincides with the helicity defined as the projection of a particle's spin along its momentum.

The left-chiral Weyl spinor was believed to describe the massless neutrinos and had been incorporated in the old Standard Model before the discovery of non-zero neutrino masses. To introduce mass terms in the Weyl equations, one must be careful about the Lorentz covariance of the equation. In order to keep the equations covariant, one should note that $m\epsilon^{-1}\psi^*$, where $\psi$ is either $\psi_L$ or $\psi_R$ and $\epsilon = i\sigma_y$, transforms like the differential part of the corresponding Weyl equation \cite{Aste}. Therefore it is possible to write the Lorentz invariant equations for massive fermions as
\begin{subequations}
\label{TCME}
\begin{align}
&i(\partial_t -\vec{\sigma}\cdot\vec{\nabla})\psi_L(x) - m\epsilon\psi_L^*(x) = 0 , \\
&i(\partial_t + \vec{\sigma}\cdot\vec{\nabla})\psi_R(x) + m\epsilon\psi_R^*(x) = 0,
\end{align}
\end{subequations}
called the left-chiral and right-chiral two-component Majorana equations. These spinors form the building blocks of higher-dimensional spinors such as the Dirac or Majorana fields. They do not correspond to any massive particles by themselves, however, and a direct observation of their dynamics is ordinarily impossible. 
In fact, Eqs.~(\ref{TCME}) is equivalent to the Majorana equation, which can be easily seen by constructing the four-component spinor $(\psi_R,\psi_L)$. In 1+1 or 2+1 D, the situation simplifies because a Majoranon can be represented by a two-component spinor and it can be shown, by explicit construction of $\gamma^\mu$ in terms of Pauli matrices, that one of the two TCMEs is equivalent to the Majorana equation.

It is also possible to build a Majorana field from either of the chiral fields. To see this, for $\psi_L$ say, note that $\epsilon \psi_L^*$ obeys the equation
\begin{align}
i(\partial_t +\vec{\sigma}\cdot\vec{\nabla})\epsilon\psi_L^*(x) + m\epsilon\epsilon\psi_L(x) = 0.
\end{align}
That is, $\epsilon \psi_L^*$ behaves like a right-chiral two-component spinor. Thus noticing that Eq.~(\ref{TCME}a) mixes the left- and right-chiral spinors, let us define a four-component spinor
\begin{align*}
\begin{pmatrix} \epsilon\psi_L^* \\ \psi_L
\end{pmatrix},
\end{align*}
which obeys the Dirac equation,
\begin{align}
\label{Eq10}
i\gamma^{\mu}\partial_{\mu}\Psi_M - m\Psi_M = 0,
\end{align}
where $\gamma^\mu$ are the Dirac matrices in the chiral representation. 

We have thus shown that a four-component spinor obeying the Dirac equation can be constructed from a two-component left-chiral spinor. Using the properties of the Dirac matrices it is possible to show that $\Psi_M$ has a special property
\begin{align}
\left(\Psi_M\right)_c = \tilde{C}(\gamma^0)^T\Psi_M^* = \Psi_M,
\end{align}
where  
\begin{align*}
\tilde{C} = \begin{pmatrix} i\sigma_y & 0  \\ 0 & -i\sigma_y
\end{pmatrix}
\end{align*}
in the chiral representation. The subscript $c$ denotes charge conjugation and the (Majorana) condition states that the spinor is invariant under charge conjugation. A Majorana particle is therefore necessarily charge neutral, which is why such a particle can be its own antiparticle. 

To build up a Dirac field from two-component spinors we need two independent left-chiral spinors, say $\psi_L$ and $\chi_L$. Then $(\epsilon\psi_L^*, \chi_L)^T$ obeys the Dirac equation, but not the Majorana condition. 

The fact that the TCME can be converted into the four-component Dirac equation suggests that a two-component Majoranon is equivalent to a four-component Majorana fermion. The resulting equation can then be simulated with a single trapped ion using the method proposed in \cite{Lamata}.  Furthermore, if we restrict the spatial dimension to 1 it is possible to write down the equation of motion in the form of interacting qubits coupled to a phonon mode, where only one qubit is coupled to the phonons. This fact has already been noticed in \cite{Casanova} through different argument and allows quantum simulation of the Majorana equation using trapped two-level ions. Our argument here can thus be seen as an alternative derivation of the result, starting from the two-component formalism. The salient point is that in 1+1 D a Majoranon is described by a two-component spinor obeying either of Eqs.~(\ref{TCME}) with 1 spatial degree of freedom and therefore it should be possible to convert this to a Hamiltonian dynamics by constructing a Majorana fermion as shown earlier. To see this explicitly, consider the Dirac equation for a Majoranon confined to move in the $z$-direction. In the Majorana representation $\gamma^0_M = \sigma_y\otimes\sigma_x$, $\gamma^3_M = i\sigma_y\otimes\sigma_y$ and $\Psi_M^* = \Psi_M$, and Eq.~(\ref{Eq10}) becomes
\begin{align}
i\partial_t\Psi_M = 1\otimes \sigma_z (i\partial_z) \Psi_M + m(\sigma_y\otimes\sigma_x)\Psi_M.
\end{align}
The Hamiltonian for this Schr\"odinger equation is equivalent to (\ref{MHam}) up to an unitary transformation and thus can be readily simulated with trapped ions as shown explicitly in the reference. 

\textbf{Decomposition of a Majoranon into two Majorana fields} -
Here, it is useful to go back to Eq.~(\ref{MHam}) briefly and discuss an interesting fact. Using the unitary operation $U = i e^{-\imath \pi \sigma_{y} / 4 } \otimes e^{-\imath \pi \sigma_{x} / 4 }$ with the basis $\Psi = U \Phi \in \mathbb{R}^{4}$, such that $\psi = M U \Phi$, the ``Majorana Hamiltonian'', $H_{M}$ in Eq.~(\ref{MHam}) becomes
\begin{eqnarray}
H &=& U^{\dagger} H_{M} U, \\
&=&  \left( 1_{2} \otimes \sigma_{x} \right) p_{x} + m (\sigma_{z} \otimes \sigma_{z} )  .
\end{eqnarray}
This Hamiltonian leads to two uncoupled Dirac equations
\begin{align}
\imath \partial_{t} \phi_{\pm} &= H_{\pm} \phi_{\pm}, \\
H_{\pm} &= \sigma_{x} p_{x} \pm m \sigma_{z},
\end{align}
with $\phi_{+} = ( \Phi^{(1)},\Phi^{(2)} )$ and $\phi_{-} = ( \Phi^{(3)},\Phi^{(4)} )$ such that $\Phi = ( \phi_{+}, \phi_{-} )$. It is interesting that $\phi_\pm$ are invariant under charge conjugation, which in 1+1 D is defined as $-i\sigma_z\sigma_y\phi_\pm^*$. The uncoupling thus suggests that a Majoranon in 1+1 D can be decomposed into two Majorana fermions.
In fact it is well-known that a field (which in 1+1 D has two components) obeying the Dirac-Majorana equation, a general relativistic field equation that has both the Dirac and the Majorana mass terms, can be broken down into two majorana fermions with different masses \cite{Aste,Cheng}. We provide a simple proof here.

To construct a general proof for the Dirac-Majorana equation, we first write it in the Majorana representation (in which charge conjugation is equal to complex conjugation): 
\begin{eqnarray}
\label{Eq17}
\imath \gamma^\mu \partial_\mu \Psi = m_M \Psi^{\ast} + m_D\Psi.
\end{eqnarray}
Decomposing the four component spinor as $\Psi = (\Psi_{+} + \imath \Psi_{-})/\sqrt{2}$ where $\Psi_{\pm}$ are the Majorana fields, i.e., $\Psi_{\pm} = \Psi_{\pm}^{\ast}$, it is readily seen that
\begin{eqnarray}
\imath \gamma^\mu \partial_\mu \Psi_{\pm} = \left( m_D \pm m_M \right) \Psi_{\pm}.
\end{eqnarray}
Therefore a Dirac-Majoranon obeying Eq.~(\ref{Eq17}) can be decomposed into two Majorana fermions obeying their respective Dirac equations. The following proposal for quantum simulation thus work not only for a Majoranon but a Dirac-Majoranon too.

\textbf{A  proposal for quantum simulation} -
The decomposition described above has an interesting consequence for quantum simulation of a Majoranon: it is possible to simulate the Majorana equation by simulating two Dirac equations with the opposite mass signs. An advantage is that only a single qubit is required, so the single-qubit addressibility and qubit-qubit interaction is not an issue anymore. The price to pay is that the method requires complete state reconstruction, including the continuous degree of freem. This is because the full information on spinors, $\phi_+$ and $\phi_-$, is required in order to calculate expectation values. Despite the caveat, the method provides a good alternative as far as observing exotic physics goes, especially for the case that only requires qubit state reconstruction, which we describe below. Note that the exotic physics of the Majorana equation does not manifest itself in the relativistic regime, in which the mass term becomes negligible and the Majorana equation reduces to the Dirac equation. Instead it is most prominent in the stationary case as discussed in \cite{Lamatanew}. 

It is also possible to simulate the TCME in 2 spatial dimensions, i.e., simulate a two-component chiral field in 2+1 D, using the fact that a Majoranon can be decomposed into two Majorana fermions. 
To see this, first note that the (right-chiral) two-component Majorana equation and the Majorana equation are equal and can be written as
\begin{align}
\label{rctcme}
i\partial_t\psi = (\sigma_xp_x + \sigma_yp_y)\psi - im\sigma_y\psi^*.
\end{align}
By decomposing $\psi = (\psi_+ + i\psi_-)/\sqrt{2}$ and assuming $\psi_\pm = -i\sigma_z\sigma_y\psi_\pm^*$, 
it is easily seen that the two Dirac equations
\begin{align}
i\partial_t\psi_\pm = -(\sigma_xp_x + \sigma_yp_y)\psi_\pm \pm m\sigma_z\psi_\pm,
\end{align}
are equivalent to Eq.~(\ref{rctcme}). The same decomposition does not work in 3+1 D because the charge conjugation operator cannot be constructed for two-component spinors.  This seems to corroborate the fact that chiral spinors are the irreducible representations of the Lorentz group, i.e., they cannot be decomposed into smaller representations. One is forced to use the four-component spinors in this case, requiring a quantum simulator in 3+1 D proposed in \cite{Lamata} for example. An interesting open question is whether two qubits instead of a single four-level system can be used to simulate the Dirac equation in 3+1 D.

\textbf{Simulation procedure} -
Thus far, we have shown that in 1+1 or 2+1 D a (two-component) chiral spinor, obeying the TCME or equivalently the Majorana equation, can be either turned into a four-component Majorana field or decomposed into two two-component Majorana fields. The latter allows quantum simulation of the Majorana equation in 1+1 (2+1) D with a single qubit and one (two) continuous degrees of freedom, given that one can reconstruct the states completely. The simulation procedure can be summarized into four steps:
$(i)$ Choose a system that simulates the Dirac equation; e.g., trapped ions \cite{Lamata,Gerritsma10}, stationary light polaritons \cite{Otterbach} or optical lattice \cite{Szpak} schemes.
$(ii)$ Prepare the initial state that you want to simulate; i.e.~write down the spinor corresponding to a particular initial state and decompose it into two spinors obeying the Majorana conditions: $\psi = \psi_+ + i\psi_-$, where $(\psi_\pm)_c = \psi_\pm$. We will show specific examples below.
$(iii)$ Time evolve the states $\psi_\pm$ according to the Dirac equation with $\pm m$, respectively.
$(iv)$ Evaluate an observable in terms of $\psi_+$ and $\psi_-$.

In 1+1 or 2+1 D, the Majorana condition $\chi_\pm = -i\sigma_z\sigma_y\chi_\pm^*$ is satisfied by the basis vectors $\chi_+ = (1,-1)/\sqrt{2}$ and $\chi_- = (i,i)/\sqrt{2}$, and one can expand any initial state in terms of these basis spinors. A given state $\psi (0)$ can be expanded as $(\psi_+(0) + i\psi_-(0))/\sqrt{2}$, in terms of the Majorana fields $\psi_\pm$ using the following relations
\begin{align}
\psi_+(0) &= \frac{1}{2}\left[\psi (0) - i\sigma_z\sigma_y\psi^*(0)\right], \\ 
\psi_-(0) &= \frac{-i}{2}\left[\psi (0) + i\sigma_z\sigma_y\psi^*(0)\right]. 
\end{align}
For example, $\psi(0) = (1,0)$ is equivalent to $\psi_+(0) = (1,-1)/\sqrt{2}$ and $\psi_-(0) = -(i,i)/\sqrt{2}$. This example corresponds to a particle at rest discussed in \cite{Lamatanew}. In this case, there is no momentum dependence and the simulation only requires qubit state reconstruction \cite{tom1}. Therefore, one can simulate the Majorana equation in the rest-frame with a single trapped ion instead of two. As explained above, this corresponds to the limiting case where the differences between the Majorana and the Dirac dynamics are most prominent.

As another example, consider a general state that distinguishes between the Majorana and the Dirac equation, i.e., a state that is not charge conjugate invariant,
\begin{align}
\psi (0) = e^{ip_0x}e^{-\tfrac{x^2}{4\Delta^2}} \begin{pmatrix} 1 \\ 1
\end{pmatrix}.
\end{align}
This state can be created in the trapped ion setup \cite{Gerritsma11}. It is decomposed into 
\begin{align}
\psi_+(0) &= \sqrt{2} ~\sin(p_0x)e^{-\tfrac{x^2}{4\Delta^2}} \chi_-, \\
\psi_-(0) &= -\sqrt{2}~\cos(p_0x)e^{-\tfrac{x^2}{4\Delta^2}} \chi_-.
\end{align}
In this case, evaluation of an observable requires complete information on $\psi_\pm$, including the spatial dependence. This means the full quantum state tomography of the qubit plus the phonon mode for trapped ions, whereas in other setups, such as stationary light polaritons or an optical lattice, it would mean a spatially resolved detection scheme.
 
\textbf{Conclusion} -
A Majoranon, i.e., a particle that obeys the Majorana equation but not the Majorana condition, can be decomposed into two Majorana fermions, i.e., particles that answer to both the Majorana equation and condition. It is therefore possible to simulate a chiral spinor with previously proposed quantum simulators of the Dirac equation. Quantum simulation in 1+1 or 2+1 dimensions requires a single qubit plus one or two continuous degrees of freedom, instead of two qubits as proposed before, with the trade-off that complete state reconstruction is required. An interesting exception is for a particle initially at rest, for which the differences between the dynamics of the Majorana and the Dirac equations are most obvious. In this latter case, only single qubit state reconstruction is needed, greatly simplifying the complexity of a possible experimental implementation of our approach.

\end{document}